%% file: ms.tex
 \documentclass[]{aastex}
 \usepackage{emulateapj5}
 \usepackage{epsf}

\shorttitle{Cosmic Rays in Galaxy Clusters}
\shortauthors{Miniati {\it et al.~}}
\slugcomment{draft of \today}

\eqsecnum

\def\eg{{\it e.g.,~}}
\def\ie{{\it i.e.~}}
\def\hinv{~$h^{-1}$}
\input{macros}

\begin{document}

\title{Cosmic Ray Protons Accelerated at Cosmological Shocks
and Their Impact on Groups and Clusters of Galaxies\altaffilmark{8}}

\author{  Francesco Miniati     \altaffilmark{1,2,3},
          Dongsu Ryu            \altaffilmark{4}, 
          Hyesung Kang          \altaffilmark{5},
      and T. W. Jones           \altaffilmark{2,6}}

\altaffiltext{1}{Max-Planck-Institut f\"ur Astrophysik,
    Karl-Schwarzschild-Str. 1, 85740, Garching, Germany}
\altaffiltext{2}{Department of Astronomy, University of Minnesota,
    Minneapolis, MN 55455}
\altaffiltext{3}{e-mail: fm@MPA-Garching.MPG.DE}
\altaffiltext{4}{Department of Astronomy \& Space Science, Chungnam National
    University, Daejeon, 305-764 Korea: ryu@canopus.chungnam.ac.kr}
\altaffiltext{5}{Department of Earth Science, Pusan National University,
    Pusan, 609-735 Korea: kang@uju.es.pusan.ac.kr}
\altaffiltext{6}{e-mail: twj@msi.umn.edu}
\altaffiltext{8}{to appear in {\it The Astrophysical Journal}}

\begin{abstract}

We investigate the production of cosmic ray (CR) 
protons at cosmological shocks
by performing, for the first time, numerical simulations of large scale
structure formation that include directly the acceleration, transport and 
energy losses of the high energy particles.
CRs are injected at shocks according to the thermal leakage model and, 
thereafter, accelerated to a power-law distribution as indicated by the 
test particle limit of the diffusive shock acceleration theory.
The evolution of the CR protons accounts for losses due to adiabatic
expansion/compression, Coulomb collisions and inelastic p-p scattering.
Our results suggest that CR protons produced at shocks formed in 
association with the process of large scale
structure formation could amount to a substantial fraction of the total 
pressure in the intra-cluster medium. Their presence 
should be easily revealed by GLAST
through detection of $\gamma$-ray flux from  the decay of $\pi^0$ 
produced in inelastic p-p collisions of such CR protons with nuclei of the 
intra-cluster gas. This measurement will allow a  
direct determination of the CR pressure contribution in the 
intra-cluster medium. We also find that the spatial distribution of
CR is typically more irregular than that of the thermal gas
because it is more influenced by the underlying distribution of shocks. 
This feature is reflected in the appearance of our $\gamma$-ray 
synthetic images. Finally, the average CR pressure distribution appears 
statistically slightly more extended than the thermal pressure.

\end{abstract}

 
\section{Introduction} \label{intro.se}

Clusters of galaxies are the largest bound objects in the universe
and prove invaluable for investigations of cosmological interests.
The statistics of cluster masses and their dynamical properties, 
including, for instance, the relative proportions of baryonic 
and non-baryonic matter, are commonly used to test basic cosmological
models \citep[\eg][and references therein]{bahb99}.
While galaxies are the most obvious constituents of clusters 
in visible light, most of the cluster mass is non-baryonic, 
and even the baryonic matter is primarily contained within 
the diffuse intra-cluster medium (ICM), rather than in galaxies.
The temperature and density distribution of the ICM gas directly
reflect the dynamical state of clusters, a topic that has received
much attention recently.
While the ICM of clusters sometimes appears relaxed, it is often 
the case that high speed flows are present,
demonstrating that cluster environments can 
be violent \citep[\eg][and references therein]{masavi99b}.

The likely existence of strong ``accretion'' shocks several Mpc's 
from cluster cores developed in the course of large-scale structure
formation has been recognized for a long time 
\citep{suze72s,bert85b,ryka97a,quibsa98}.
Such shocks are responsible for the heating of the ICM up to 
temperatures of order $10^7 - 10^8$K.
However, cosmic structure formation simulations have demonstrated,
in addition to accretion shocks and discrete merger shocks,
the existence of somewhat weaker shocks ``internal'' to the ICM
that are very common and complex 
\citep{minetal00}.
Since clusters tend to form at the intersections of cosmic filaments,
they accrete matter in unsteady and non-isotropic patterns
through large scale flows propagating down filaments and producing shocks
as they impact the ICM. 
When cluster mergers take place, the accretion shocks
associated with the individual clusters add to the shocks
that form in direct response to the merger.
The net result of all of this is a rich web of relatively weak shocks, which often
penetrate into the inner regions of the clusters \citep{minetal00}.
Shocks resulting from discrete cluster merger events have already been 
identified by the observation of temperature structure 
in the ICM \citep[\eg][and references therein]{masavi99b}. Such shocks have also 
been claimed as the acceleration sites for relativistic electrons 
responsible for the non-thermal emission observed from clusters
in the radio, \hxr and \euv 
\citep[\eg][and references therein]{tana00,ebkk98,robust99}.

Magnetic fields are commonly observed in 
the large scale structures \cite[\eg][]{kronberg94}.
They may have been seeded at shocks in the course 
of structure formation and amplified up to $\mu$G 
level in clusters and, perhaps, also in filaments and super-clusters
\citep{kcor97,rykabi98}.
Because shock waves in the presence of even modest magnetic fields are sites of
efficient cosmic ray (CR) acceleration \citep[\eg][]{drury83}, 
structure formation might imply copious generation of high energy 
particles, including both protons and electrons.
In fact, according to diffusive shock acceleration theory \citep{drury83},
as much as several tens of percent of the 
kinetic energy of the bulk flow 
associated with the shock can be converted into CR protons 
\citep{eichler79,beel99}.
Given their huge size and long durability, large scale structure shocks
have also been suggested as possible sources of the very high energy 
CR protons up to a few $\times 10^{19}$ eV \citep{karyjo96,karabi97}.

The overall energetics of the ``cosmic'' (accretion and internal)
shocks is generally consistent with the production of CRs containing
a significant energy fraction.
Typical flow speeds in and around clusters will be
$v_f \sim (2G~M_{cl}/R_{cl})^{1/2} \sim 2\times 10^3$
km s$^{-1}$, leading to an available power for CR production
at accretion shocks $\Phi_E \sim
\rho_{b} v^3_f R^2_{cl} \sim 10^{46}$ erg s$^{-1}$
using $M_{cl} \sim 10^{15} M_{\sun}$ and $R_{cl} \sim 2$ Mpc.
According to \citet{minetal00}, where the statistics of cosmic shocks
in both SCDM and $\Lambda$CDM cosmologies are explored, 
accretion shocks appear to be less important as potential sources of
CRs than internal shocks, despite the typically greater strength of the accretion
shocks.
The reason is that internal shocks repeatedly process the ICM
material, whereas the accretion shocks do it only once
with low density background material
\cite[\cf][for $\Phi_E$ in clusters estimated from simulations]{minetal00}.

Observations of radio emission from CR electrons 
as well as radiation excess in the hard X-ray and 
possibly EUV bands, 
have recently stimulated much discussion about 
cluster physics \citep[\eg][and references therein]{sarazin99}.
They have provided information regarding the energy density of CR electrons.
CR protons produce $\gamma$-rays through $\pi^0$ decay following inelastic
collisions with gas nuclei.
While such $\gamma$-rays have not yet been detected from clusters
\citep{sreeku96}, recent estimates have shown that $\gamma$-ray
fluxes from the nearest rich clusters, such as Coma, are within the
range of what may be detected by the next generation of $\gamma$-ray
observatories \citep{cobl98,blasi99,doen00}. Their detection will provide
essential information about 
the presence of and amount of energy carried by
CR protons in the ICM (\cf \S \ref{grf.se}). 

Relativistic protons below the ``GZK'' energy threshold for
photo-pion production due to interaction with the cosmic
microwave background photons
(\ie $E\lesssim 10^{9.5}$ GeV) do not suffer significant energy losses
in cluster environments during a Hubble time \citep{bbp97}. 
In addition, up to somewhat lower energies ($\sim 10^6$GeV), even conservative
estimates of diffusion rates would confine CRs within clusters since their
formation \citep{voatbr96,bbp97}.
Therefore, CR protons, once introduced, should accumulate in clusters,
with the possibility of impacting on a wide range of issues. 
Some topics that could be impacted include
cluster formation and evolution, as well as cluster mass estimates
based on the assumption of ICM Hydrostatic equilibrium.
The dynamics of cooling flows also would obviously be affected.

The analysis of \citet{minetal00} showed that the most common shocks
in the ICM have typical Mach numbers less than
10, with a peak around $M \sim 4-5$. That is significant, because such
shocks are strong enough to transfer as much as $20-30$\% of the bulk 
kinetic energy into CR pressure, but are only mildly modified
by the CR back-reaction \citep{jmrk00}. 
Thus, the test-particle approximation, in which such dynamical feed-back
is ignored, should be a 
reasonable, physically justified  assumption to begin investigating the production of CR
at cosmic shocks.

In this paper we investigate the acceleration of CR protons at cosmological
shocks by means of numerical calculations. For the first time the CR
population is {\it directly} included in the computation with 
particle injection, acceleration and energy losses
calculated in accord with the properties of the local enviroment
in which the particles are propagating.
Here, our focus is CR {\it protons}, while CR {\it electrons} 
will be discussed in a companion paper \citep{mjkr01}. 
There are additional sources of CRs in clusters, of course, such as active
galaxies \citep{ebkw97,bbp97}. We do not attempt to include them in our
current simulations, since our goal is to understand the role of
structure shocks. However, we do call attention in our discussions
to some expected
differences between shock CR sources and point sources as appropriate.
The results of our modeling efforts should provide some initial
clue as to how these different sources can be distinguished observationally.

The paper is organized as follows. In \S \ref{nusi.se} we 
outline the numerical methods adopted for our study.
In \S \ref{res.se} the results are presented.
A discussion on the implication of the results of this paper
is given in \S \ref{disc.se}, whereas the main conclusions 
are summarized in \S \ref{concl.se}.

\section{Numerical Simulations} \label{nusi.se}

\subsection{Cosmological Hydrodynamic Simulations}

For the numerical calculations we employed 
an Eulerian ``TVD'' hydro~+~N-body cosmological
code \citep{rokc93}. 
Since the computation involves a new quantity never simulated
before in this context, \ie CRs, we have decided to begin 
the study from the simpler case of the standard cold dark matter (SCDM) model,
leaving the currently more favored CDM + $\Lambda $ model 
as the natural follow-up step for future work.
Although it is well known that SCDM is not a viable model anymore 
\citep[\eg][]{ostrk93}, we have chosen the key cosmological parameters
so that properties of the simulated collapsed objects are consistent with 
observations, thus allowing assessments of their general 
characteristics.
For instance, we adopted rms density fluctuations on a scale
of 8\hinv Mpc  to be defined by $\sigma_8 = 0.6$,
which is incompatible with COBE results and the SCDM universe, 
yet induces the emergence of a reasonable population of collapsed objects 
in simulations of large scale structure formation \citep{osce96}.
We have also adopted the following key parameters:
spectral index for the initial power spectrum 
of perturbations $n$ = 1, normalized Hubble constant 
$h \equiv$ H$_0$/(100 km s$^{-1}$ Mpc$^{-1}$) = 0.5, 
total mass density $\Omega_M = 1$, and baryonic fraction $\Omega_b = 0.13$. 
In addition, we use a standard metal composition with hydrogen and 
Helium mass fractions
$f_H = 0.76 $ and $f_{He}=0.24$ respectively. Thus,
for a fully ionized gas the mean molecular weight 
used in the temperature definition is $\bar\mu =0.59$.

In order to simulate a cosmological volume large enough to contain
groups/clusters with a sufficient resolution,
we select a cubic comoving region of size 50 \hinv Mpc and use 
$256^3$ cells for baryonic matter and $128^3$ dark matter (DM) particles.
This corresponds to a spatial resolution of $\sim$200\hinv kpc.
A few comments about the effects due to finite numerical resolution are 
appropriate here in order to define the scope of our findings. 
In general a coarse grid limits the structures
that can form during the evolution of the simulated systems.
This implies first that density peaks are smoothed out while
masses of the structures are conserved.
As a consequence, quantities such as the X-ray and $\gamma$-ray
luminosity, which depend on the square of the density,
will be reduced. 
The effect is stronger for lower temperature groups/clusters which have similar 
structures as the larger clusters but smaller physical scales.
Thus, because of resolution effects, these types of emissions (X-ray, $\gamma$-ray)
will be systematically underestimated and will lead to steeper intra-cluster 
temperature dependences in our numerical calculation.
Previous numerical studies carried out to test the performance
of the hydrodynamic part of the code employed here indicate that 
the X-ray thermal emission ($\propto n_{gas}^2$)
is underestimated by a factor of a few \citep{ceos99b}. 
Secondly, although shocks are captured cleanly within only a few 
computational zones, that still amounts to a fair fraction of 
the cluster size. This introduces an uncertainty in the location of
the shocks and reduces both the shock surface extension 
and complexity. However, the total flux of kinetic energy through shocks
should not be affected, as indirectly attested by the 
fact that the computed intra-cluster temperatures are quite accurate 
\citep{kangetal94, frenketal99}.
Thus we expect our results to be physically correct, although
further work is required in order to achieve high quantitative accuracy.
Since this is the first attempt to investigate such a problem,
the level of accuracy characterizing our simulation should
be sufficient enough to explore qualitatively
the physical impact that CRs may have in cosmological environment
and to provide a preliminary assessment of their observability.

\subsection{Cosmic Ray Injection and Acceleration} \label{injac.se}

The evolution of the CR population in the simulation
is computed via passive quantities 
by the code COSMOCR \citep{min01}. In the following sections 
we provide a brief description of the physical processes included 
in this code, \ie CR injection at shocks and spatial 
transport and energy losses.

In the calculation
CRs are injected at shocks according to the ``thermal leakage''  model
\citep[\eg][]{elei84,kajo95}.
In this model the post-shock gas is assumed to have mostly thermalized to
a Maxwellian distribution, $f(p)_{\rm Maxwell}$, 
characterized by the downstream temperature, $T_{\rm shock}$. 
Thermal protons in the high energy tail of such a Maxwellian distribution
can escape back upstream of the shock if their speeds are sufficient to
allow them to
avoid being trapped by the plasma waves that moderate the shock 
\citep{mavo95}.
Those protons are injected into the diffusive shock acceleration mechanism 
and can be accelerated to high energies.
In the present calculation, the momentum threshold 
for injection, $p_{inj}$, is set to a few times the peak thermal value, \ie
\begin{equation}
p_{inj} = c_1\,2\sqrt{m_p k_{\rm B} T_{\rm shock}} 
\end{equation}
where $m_p$ is the proton mass, $k_{\rm B}$ is the Boltzmann's constant,
$T_{\rm shock}$ is the post-shock gas temperature,
and $c_1$ is a parameter which
regulates the number of injected particles (see below).
This limit was chosen to be consistent with more detailed, nonlinear CR acceleration
simulation results described at the end of this subsection.
In the test-particle limit adopted here, the diffusively accelerated 
CRs emerging from a shock are characterized by a power law 
distribution function given by 
\begin{equation}
f(p)_{\rm shock}= f(p_{inj})_{\rm Maxwell} \left(\frac{p}{p_{inj}}\right)^{-q} .
\end{equation}
extending from $p_{inj}$ to $p_{max}$. Here, the log-slope 
is determined by the shock strength, \ie $q=3r/(r-1)$ 
(where $r$ is the shock compression ratio);
and the normalization is given by the value of
the Maxwellian gas distribution at momentum $p_{inj}$. Thus the
thermal distribution, $f(p)_{\rm Maxwell}$, and CR distributions, 
$f(p)_{\rm shock}$, join smoothly in terms of the momentum
coordinate. This power-law CR distribution is assigned to each
grid cell that is identified as ``being shocked'' within a time
step in the numerical simulation.
The physical upper bound to the CR momentum distribution is 
determined by several factors, including the time available for
acceleration compared to the mean time for particles to re-cross
the shock due to the competition between wave scattering and advection,
the extent of a shock  compared to particle scattering lengths and 
any competition from energy losses during acceleration.
For parameters appropriate to groups/clusters we expect the acceleration
to proceed relatively quickly up to momenta at least as great as $10^6$GeV/c.
CRs with even higher energy can be produced in principle \citep{karyjo96}. 
Conservative estimates indicate that these very high energy CRs can 
diffuse out of clusters carrying away some energy. That would affect 
our results only if the spectra of the accelerated CRs are significantly 
flattened with respect to the test particle limit above our adopted
momentum upper limit. However, this type of behavior 
typically is expected only for CR dominated and strongly modified shocks.
From the observed properties of the intercluster medium,
where most of the pressure is thermal, 
most likely that type of shock does not occur.

In the simulation we assume the power law is formed within one
dynamical time step up to $p_{max} = 10^6 {\rm GeV}/c$,
so that spatial diffusion of CRs can be neglected and the 
computational cost much reduced. 
To follow the evolution of the
CR distribution in detail from injection to $p_{max}$
would be completely impractical, since it would necessitate numerical
resolution on the scale of the physical thickness of the shocks \citep{jre99}.
Similarly, since spatial diffusion of CRs away from shocks is likely
to be slow below $p_{max}$, it is neglected there, but 
COSMOCR does include adiabatic energy changes, as well as the
energy losses from Coulomb and inelastic p-p collisions with the thermal ICM.
To do this a Fokker-Planck equation is solved that has been
integrated over finite momentum bins to take advantage of the
near-power-law form of the CR momentum distribution, $f(p)$.
In effect, the momentum space is divided into 
8 logarithmically equidistant intervals, bounded by 
$ p_1, ...p_8, p_9= p_{max}$, which we refer to here as momentum bins.
Within each momentum bin, $j$, we assume 
$f({\bf x}_i,p) \propto p^{-q_j({\bf x}_i)}$,
where $q_j({\bf x}_i))$ is determined self-consistently 
from $n({\bf x}_i,p_j)$ defined below
and the required continuity of $f(p)$.
At each computational spatial grid point, ${\bf x}_i$, and for 
each momentum bin, $j$,
we define the number density as 
\begin{equation}
n({\bf x}_i,p_{j})= \frac{4\pi}{3}\;
\int_{p_{j}}^{p_{j+1}} f({\bf x}_i,p) p^2 dp.
\end{equation}
For a full description of the code COSMOCR we refer to
\citet[][but see also \citealt{jre99}]{min01}.

Before concluding this section, we return for a moment to the ``injection''
parameter $c_1$ which deserves some further comments.
As already pointed out the value of $c_1$ determines 
the fraction $\eta_{inj}$ of particles in the post-shock 
gas with density $n_2$ that are injected into the CRs as follows 
\citep{min01}:
\begin{equation}
\eta_{inj} \equiv \frac{n_{inj}}{ n_2}
= 8\,\sqrt{\frac{2}{\pi}} \, c_1^3\,e^{-2c_1^2}
\, \frac{\left(\frac{p_{max}}{p_{inj}}\right)^{3-q}-1}{3-q}
\end{equation}
where, obviously, $n_{inj}$ is the number of injected particles.
In addition, 
we find the ratio of the CR pressure
to ram pressure ($\rho_1 \,u_s^2$, which 
supplies the energy for the cosmic rays) to be \citep{min01}
\begin{equation} \label{pcram.eq}
\frac{P_{cr}}{\rho_1 u_s^2} =
\frac{8}{3}\sqrt{\frac{2}{\pi}} \, c_1^3\,e^{-2c_1^2}
\,\left(\frac{m_pc}{p_{inj}}\right)^{3-q}
\, \left(\frac{c}{u_s}\right)^2 \,
\frac{\left(\frac{p_{max}}{p_{inj}}\right)^{4-q}-1}{4-q}
\end{equation}
where $c$ and $u_s$ are the light and shock speed, respectively
and where we neglected the non-relativistic contribution of the 
CR pressure (this is justified since $q\simeq 4$ so that
most of the CR pressure is produced by relativistic particles; see \S \ref{res.se}).
Both observational and theoretical studies of diffusive shock 
acceleration suggest that canonical values of $c_1$ should be around
$2.3-2.5$, corresponding to the value of $\eta_{inj}$ ranging between
a few $\times 10^{-3}$ to $10^{-4}$ \citep{lee82,quest88,kajo95}.
According to the simulations by \citet{gijoka00} where a self-consistent
injection treatment based on the plasma physical model of \citet{mavo95} is
adopted, the above injection parameter is in fact very reasonable.
For the flow parameters relevant for the cosmic shocks in our simulations, 
we find that a value of $c_1=2.6$ produces an injection efficiency 
$\eta_{inj}$ and a post-shock CR pressure consistent with the saturation value
obtained from numerical
studies of shock acceleration in which the back reaction of
the particles is accounted for \citep{beksye95}.
Note that, for this reason, our value of $c_1$ is somewhat larger than the canonical
values due to the test-particle treatment.
For a shock with Mach number $M\sim 4$ and speed $u_s \sim 10^3$km s$^{-1}$,
$q\sim 4.2$ and we evaluate (for $p_{max} =10^{15}$ eV)
\begin{equation} \label{pcramval.eq}
\frac{P_{cr}}{\rho_1 u_s^2} \nonumber
 \sim  
9.5 \times 10^{-2} \left(\frac{c_1}{2.6}\right)^{4.2}\,
e^{-2(c_1^2-2.6^2)} \, 
\left(\frac{M}{4}\right)^{-1.2}\,
\left(\frac{u_s}{10^3\mbox{km s}^{-1}}\right)^{0.8}
\end{equation}
Since most of the flow kinetic energy is processed by cosmic shocks with
$M\sim 4-5 $, we expect from experience with detailed CR shock simulations that
up to 10-30 \% of it will be converted 
into CR energy with the above value of $c_1$.
Detailed simulations also show that modifications to such shocks are
small enough that the form of the CR spectrum is not substantially
changed from the test-particle theory \citep{jmrk00}.
So our choice of $c_1$ is also consistent with our assumption that
the CR acceleration can be treated by the test-particle theory. 

\subsection{Extracting Global Properties of Groups/Clusters} \label{glpr.se}

After the calculation was completed, the simulated groups/clusters have
been identified by the DM-based ``spherical over-density'' method 
described in \citep{laco94}.
The details of the group/cluster identification procedure can be found 
in \citet{minetal00}.
Global group/cluster properties, such as core temperature, average pressure,
emissivity at various wavelengths etc., were calculated
by averaging or integrating the quantities over the group/cluster volume. 
These global properties have then been studied by means of correlation
plots in order to make predictions about the quantities under 
investigation (and often yet to be measured) in terms of the
well established ones (see \S 3.1-3.3). 
We point out from the outset that, because of the relatively small 
computational box, the temperature of the simulated collapsed
objects only ranges between 0.3 and 3 keV. Nevertheless,
after determining the temperature dependence of the various
properties of interest we extrapolate their values beyond these 
temperature limits and make estimates even for rich 
clusters such as Coma which are easier to observe. 
So long as there are no important scales involved, these extrapolations should be reliable.

Two-dimensional projections of individual group/cluster
structures have also been constructed from 
the data-set, either as slices through the simulated volume
or as synthetic images of cluster emissions. The synthetic images,
computed here in the X-ray and $\gamma$-ray bands, 
are produced by means of a projection code 
 \citep{iant01} that
integrates the ``emissivity'' along the line of sight in the 
optically thin plasma approximation.
These images allow a more in-depth inspection of the spatial distribution
of the quantities of interests, but for space reasons are limited
here to only a few examples (see \S 3.4).

In general, the $\gamma$-ray flux and the surface brightness 
for the synthetic images have been calculated by
arbitrarily 
setting the groups/clusters to a luminosity distance of 
about 70 \hinv Mpc (\ie $z=0.023$)
corresponding to the Coma cluster (Abell 1656).
Since our grid resolution amounts to $\sim 200$ \hinv kpc,
at this distance the minimum size of a pixel 
of the synthetic image corresponds to 9.8 arc-min square. 

\section{Results}  \label{res.se}

\subsection{Cosmic Ray Energy Content}

The CR pressure at ${\bf x}_i$ is defined by
\begin{equation}
P_{cr} ({\bf x}_i) = \frac{4\pi}{3}\; c \,\int_{p_{inj}}^{p_{max}}
f({\bf x}_i,p)\,\frac{p^4}{(m_p^2 c^2+p^2)^{1/2}}\;dp
\end{equation}
with $f({\bf x}_i,p)$ reconstructed from $n({\bf x}_i,p)$
and $q({\bf x}_i,p)$ as described in \S2.2.
From our simulations we find that $n({\bf x}_i,p)$ has a strong
spatial dependence, whereas $q({\bf x}_i,p)$ assumes a relatively 
narrow range of values, mostly between 4.01 and 4.2. 
The thermal pressure 
obeys by the equation of state for an ideal gas 
\begin{equation}
P_{th} ({\bf x}_i)=n_{tot} ({\bf x}_i)\,k_B\,T  ({\bf x}_i),
\end{equation}
where $n_{tot}=n_{ion}+n_e=(2f_H+3f_{He}/4) \rho_{gas}/m_p$ is the gas 
number density inclusive of both ions and electrons, and $T$ 
the gas temperature.
From the thermal and CR pressures defined at each cell,
we calculate the mean thermal and CR pressure of groups/clusters within a 
sphere of radius $R_{cl}\simeq 0.5$ \hinv Mpc from the cluster center as
\begin{equation}
\left( \begin{array}{l}  P_{th} \\ 
P_{cr} \end{array} \right)_{cl}
=  \frac{1}{\sum_i w_i} \; \sum_i w_i\,
\left( \begin{array}{l} P_{th}({\bf x}_i) \\ 
P_{cr}({\bf x}_i) \end{array}\right)
\end{equation}
where the summation over $i$ extends to the groups/clusters
volume $V=4 \pi\,R^3_{cl}/3$ and 
the weight function $w_i$ is given by the portion of 
each computational cell within $V$.
Given our resolution, the volume within a radius of $0.5$ \hinv Mpc 
typically includes about 65 computational cells.
We also calculate the groups/clusters core temperature, $T_x$ as a volume
averaged ICM temperature within the same volume.

\begin{figure}
\plotone{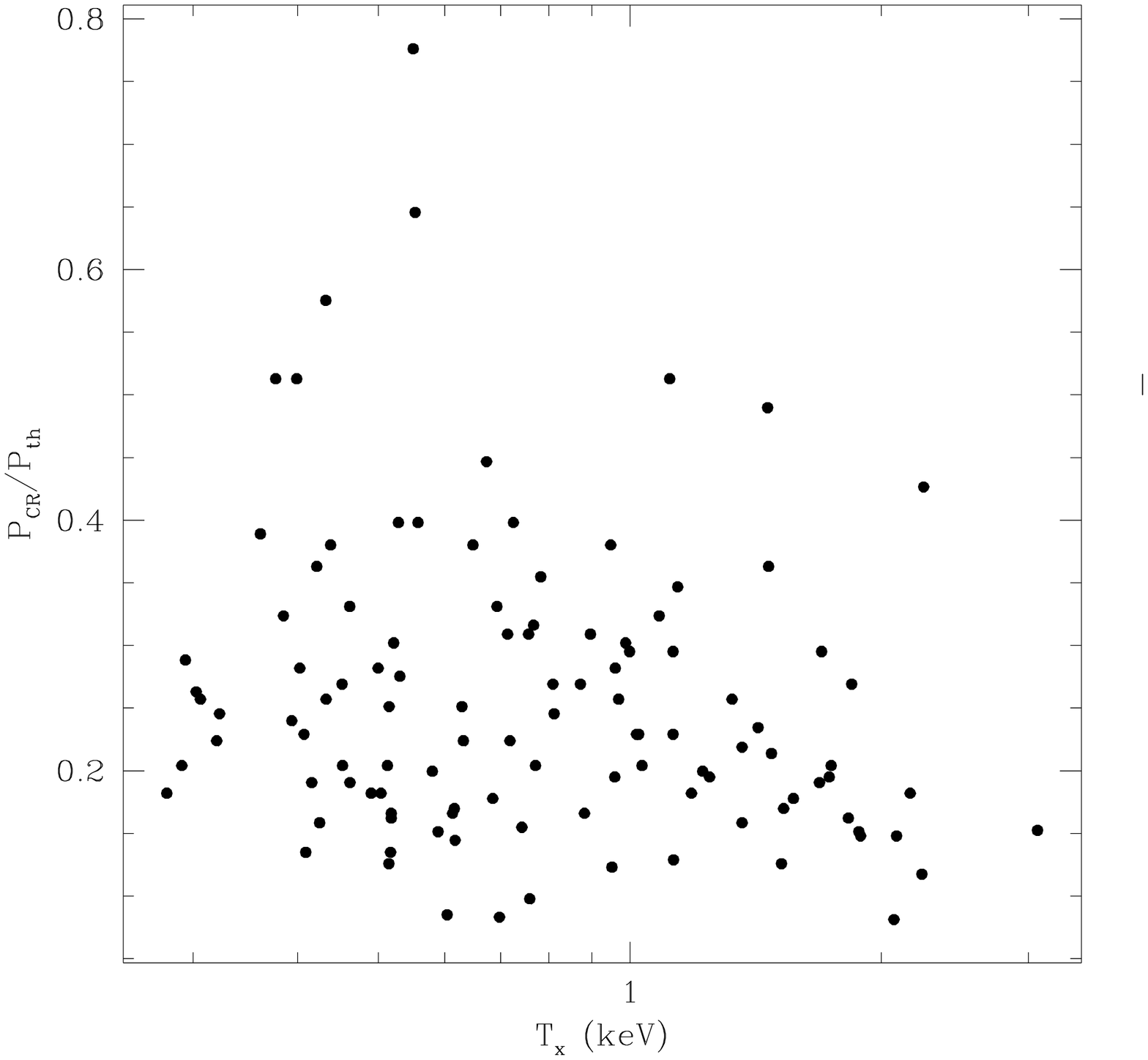}
\figcaption{Ratio of CR to thermal pressure averaged over
the group/cluster volume within 0.5 \hinv Mpc plotted as a function 
of group/cluster core temperature. \label{pctvol.fig}}
\end{figure}

The first important result of this study is illustrated in 
Fig. \ref{pctvol.fig}. 
There we plot the ratio $(P_{cr}/P_{th})_{cl}$
for groups/clusters at the current epoch (\ie $z=0$)
as a function of the core temperature, $T_x$. 
The values of this ratio, 
$(P_{cr}/P_{th})_{cl} \simeq (E_{cr}/2E_{th})_{cl}$,
where $E$ stands for energy density, offers a first-order 
indication of the relative importance of the two components for the
dynamics of groups/clusters. From Fig. \ref{pctvol.fig}, we 
can read that a significant fraction (up to $\sim 45$\%)
of the total pressure inside today's groups/clusters could be borne by CRs.
We note here that the actual content of CR pressure (and
energy density) depends on the injection parameter for which we have made
an educated estimate based on published studies related to
the theory of diffusive shock acceleration. 
As already pointed out in \S\S \ref{intro.se} and 
\ref{injac.se} here we are only allowed a simplified 
treatment of the injection mechanism at shocks.
Therefore, the result in Fig. \ref{pctvol.fig} should not 
be interpreted as a precise estimate of the CR 
content inside groups/clusters of galaxies. Rather, it provides  
a qualitative, yet sound, insight that 
CRs might be quite important for the dynamics of those objects.
Considering the difficulties in following the physics of CR acceleration
self-consistently
in multi-dimensional simulations, the quantitative estimate of the
total CR content needs to be done through measurements
of $\gamma$-ray fluxes from groups/clusters, as we shall see below.

The ratio of pressures plotted in Fig. \ref{pctvol.fig} 
does not show any particular trend except for a slight 
reduction toward higher temperatures. This 
might be due to our injection model which,
according to eq. \ref{pcram.eq}, produces a 
higher ratio of CR to thermal pressure for shocks
with smaller velocities ($u_s$) and similar Mach numbers (yielding similar $q(M)$); i.e., cooler preshock gas.
It is possible that such shocks occur in cooler groups/clusters,
characterized by lower accretion velocities and similar pre-shock temperatures.
However, the trend in Fig. \ref{pctvol.fig} 
could also be due to adiabatic compression inside the cluster, which increases 
the thermal pressure at a higher rate than the CR one.
At least we can be sure that 
the apparent scatter there is in part a reflection of
the diverse CR acceleration histories of groups/clusters.
In addition, part of this scatter can also be due to the
different spatial distribution of the thermal and
CR components  as we shall see in \S \ref{spdis.se} and \ref{xgaim.se}.

\subsection{Spatial Distribution of Thermal and Cosmic Ray Pressure} 
\label{spdis.se}

Another feature of interest is the distributions
of thermal and CR pressures inside groups/clusters.
The difference in the distributions of the two pressure components
is an important detail, because, according to the condition for
hydrostatic equilibrium,
\begin{equation}
{dP_{tot}(r) \over dr} = -{G M_{cl}(r) \rho_{gas}(r) \over r^2},
\end{equation}
so, it is the total pressure gradient that responds to the group/cluster 
mass enclosed in a volume of radius $r$. 
Thus the spatial distribution (gradient) of $P_{cr}$ is as 
important as the amount of CR energy content 
($E_{cr} \simeq 3\,P_{cr}$) itself, once $P_{cr}$ becomes
dynamically significant.

\begin{figure}
\plotone{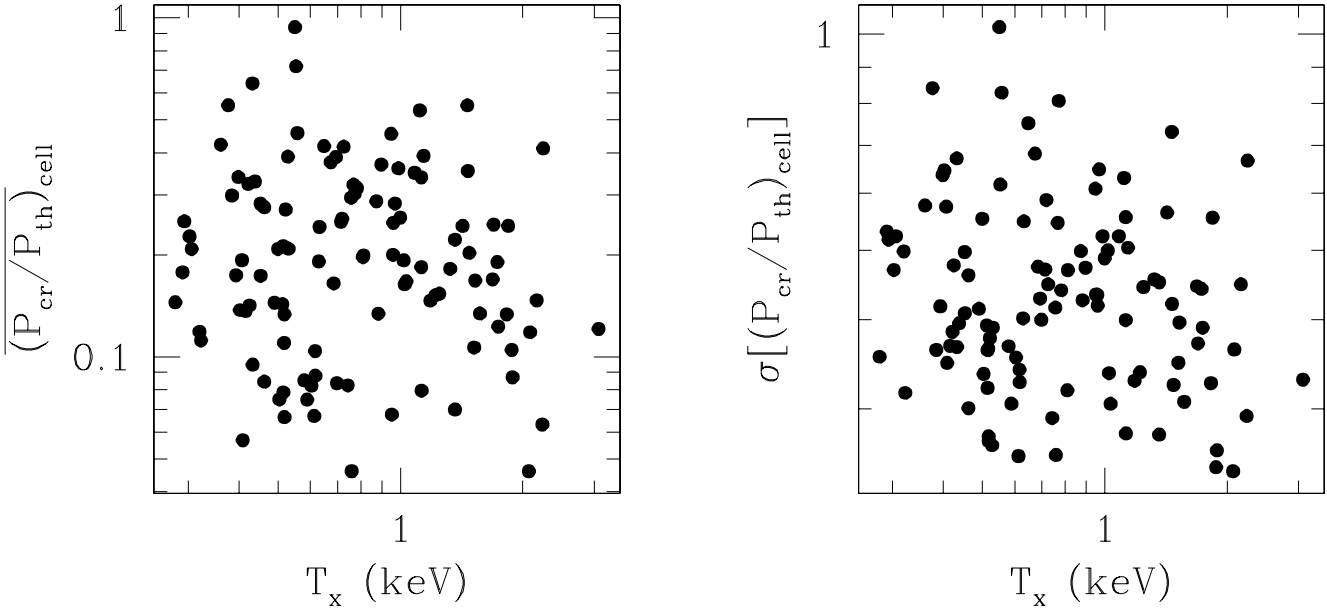}
\figcaption{Left panel: average of the cell by cell evaluation of
 $(P_{cr}/P_{th})_{cell}$. Right panel: standard deviation.
\label{ppcav.fig}}
\end{figure}

First we consider the ratio of CR to thermal pressure evaluated within
each computational cell inside groups/clusters; \ie $(P_{cr}/P_{th})_{cell}$. 
The average of this quantity over the cells inside each group/cluster volume
is plotted in the left panel of Fig. \ref{ppcav.fig}, 
as a function of the group/cluster temperature.
This is similar, but not identical to $(P_{cr}/P_{th})_{cl}$ in
Fig. \ref{pctvol.fig},
The standard deviation of $(P_{cr}/P_{th})_{cell}$
values within each group/cluster is shown in similar fashion 
on the right panel of the same figure.
It is clear that the dispersion around the average value is as large
as the average itself, indicating a strong variation 
of $(P_{cr}/P_{th})_{cell}$ inside each group/cluster volume.
Note that the gas temperature and the slope of the CR distribution,
$q$, are approximately uniform within groups/clusters.
So, this pressure behavior should be a reflection of the different spatial 
distributions of gas and CRs in the simulation.

\begin{figure}
\plotone{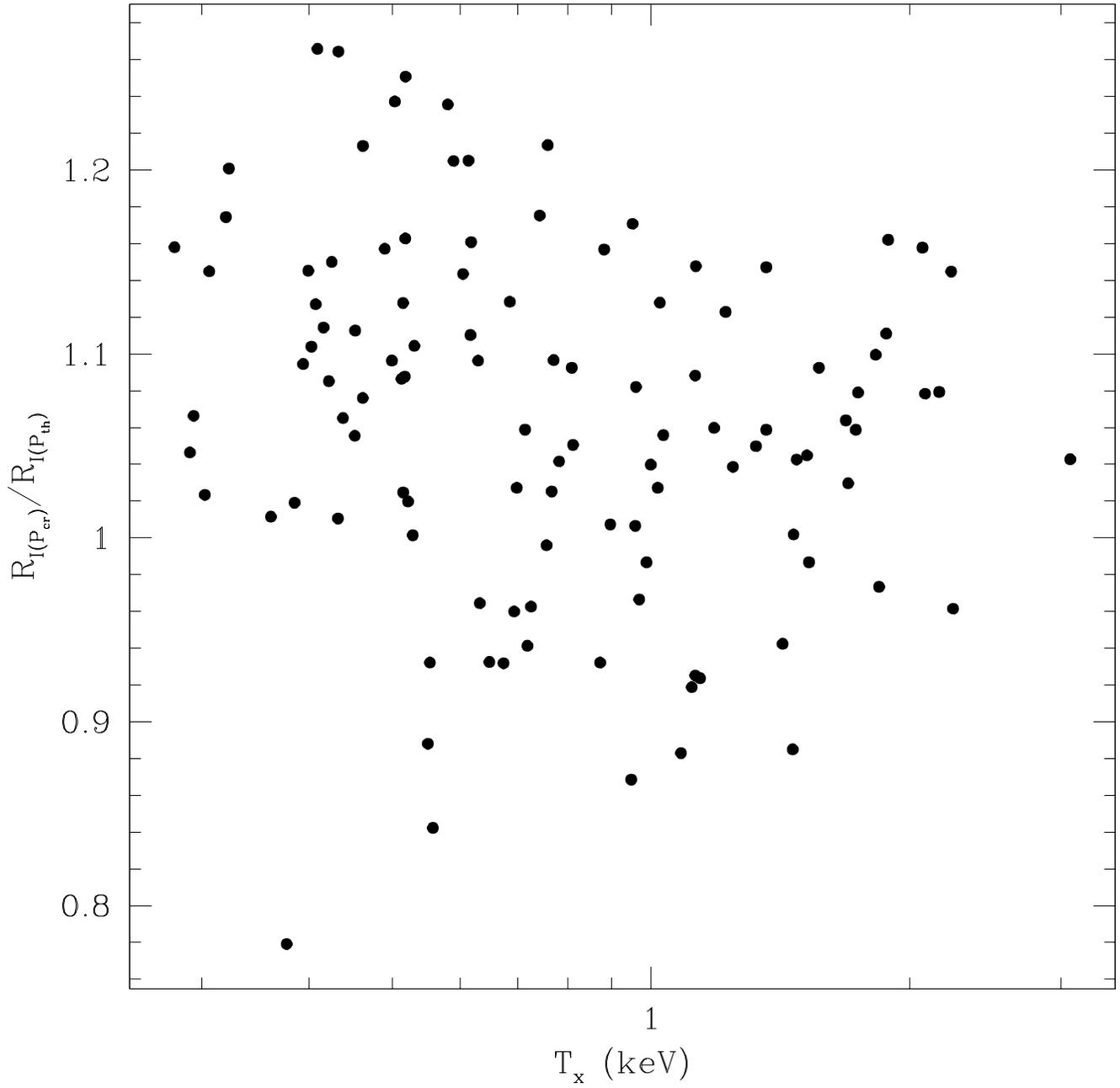}
\figcaption{Ratio of pressure-weighted rms radii of CR to
thermal pressure defined in Eq. (3.5). \label{ppim.fig}}
\end{figure}

In order to quantify the difference in spatial distributions of the two 
pressures, we define a pressure-weighted mean square radius
\begin{equation}
R_I^2 = \frac{\sum_i P_i r_i^2}{\sum_i P_i}
\end{equation}
where $P_i$ and $r_i$ indicate the pressure
(either thermal or CR) and the distance from the group/cluster center 
of the $i$-th computational cell, respectively.
In Fig. \ref{ppim.fig} we plot the ratio of $R_I$ relative to the CR
and thermal pressure, \ie $R_{I(p_{cr})}/R_{I(p_{th})}$. The plot shows that 
this ratio is close to one, with a marked tendency to values slightly
larger than one. That is, CR pressure would be 
distributed more diffusely than gas pressure in groups/clusters.
Caution is needed here, since the diameter of the collapsed objects
covers only about 5 computational cells.
Thus, although the collapsed objects have been formed with adequate resolution
to assure their basic properties, the fine 
details of their structures have not been captured. Nevertheless,
the systematic difference could be connected to 
the mechanism of CR production.
In fact, strong shocks in the simulations
are more commonly located at the outskirts of the collapsed object.
There the ratio of CR to thermal pressure is therefore higher,
causing $R_{I(p_{cr})}$  to be slightly larger than $R_{I(p_{th})}$.

\subsection{Gamma Ray Flux}  \label{grf.se}

A direct observational consequence of CR protons in groups/clusters is the
$\gamma$-ray emission from $\pi^0$ decay.
The $\gamma$-ray emissivity was calculated in each 
cell from the gas density and the CR proton distribution. The
cross sections were computed according to the GALPROP routines
\citep{most98}. The number of $\pi^0$ produced in each hadronic 
interaction, $\xi_{\pi^0}$, 
increases rather slowly with the proton energy,
$E_p$,
roughly as $\xi_{\pi^0}\simeq [(E_p - E_{ths})/\mbox{GeV}]^{1/4}$ for $E_{ths} 
\lesssim E_p \lesssim 10^4$ GeV ($E_{ths}= 1.22$ GeV is the 
energy threshold of the process; see, \eg
\citealt{masc94}). For a proton power-law
distribution with kinetic energy, $f_{cr}(T_p) \propto T_p^{-q+2}$,
CRs at energy $T_p$ generate
a number of $\pi^0$ which roughly scales as 
$j_{\pi^0} \propto (T_p/\mbox{GeV})^{-(q-5/2)}$. Then,
since $q \gtrsim 4$ the majority of the integrated $\gamma$-ray flux
is contributed by CR protons in the low energy component. 

\begin{figure}
\plotone{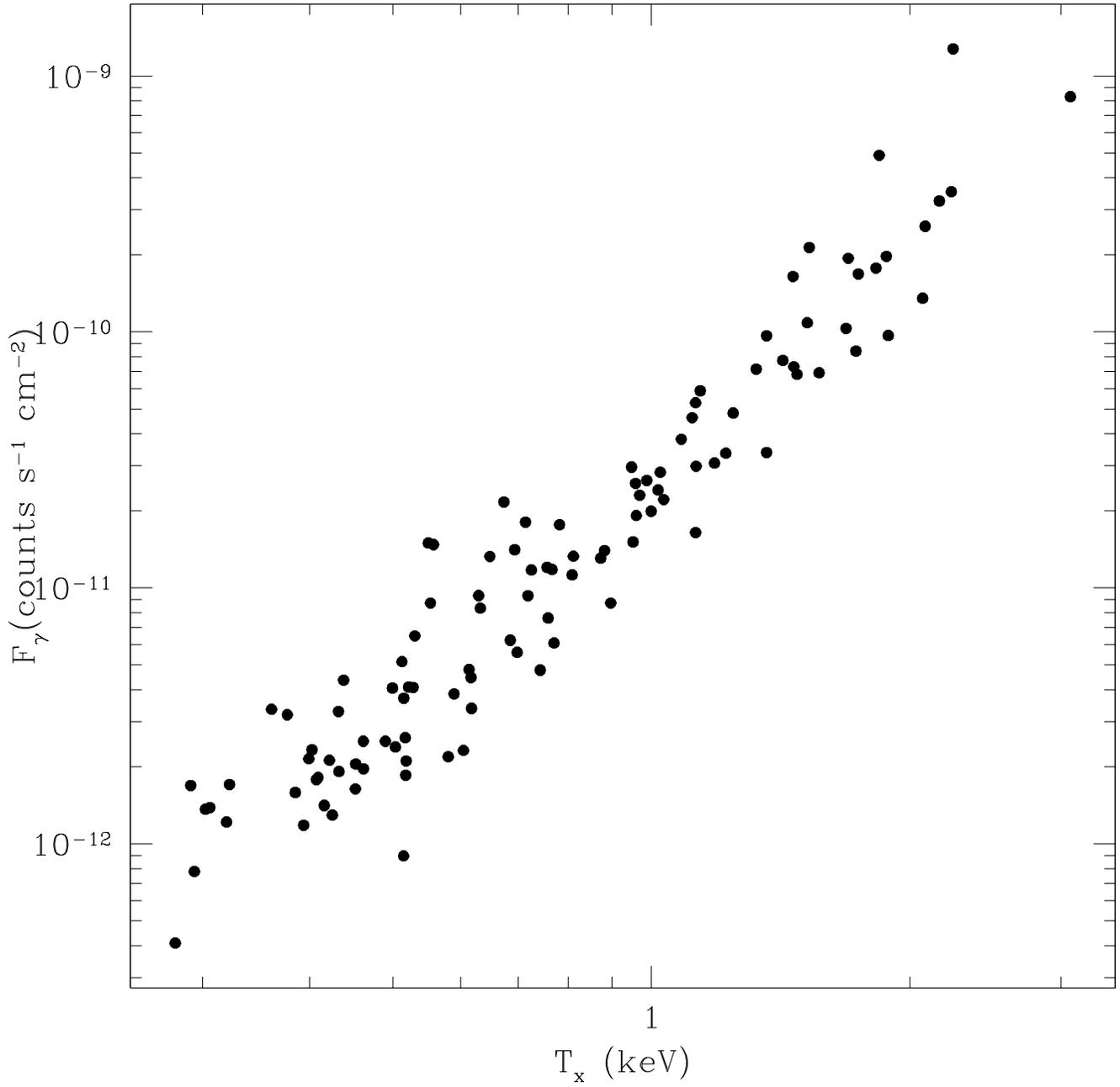}
\figcaption{$\gamma$-ray flux as a function
of group/cluster core temperature. \label{fga.fig}}
\end{figure}

In Fig. \ref{fga.fig} we report the expected $\gamma$-ray flux
above 100 MeV, $F_{\gamma}$, from a volume within 1.3 \hinv Mpc from the group/cluster center, 
as a function of the core temperature, $T_x$.
An integration volume larger than a typical group/cluster core region
of 0.5\hinv Mpc is chosen, because the CR proton distribution extends
out further to where the accretion shocks are found.
When fitting the $F_{\gamma}-T_x$ relation with a power law curve 
from a simple $\chi^2$ analysis we get
\begin{equation}
F_\gamma = 7.4\times 10^{-9}\; 
\left(\frac{T_x}{6.72\mbox{keV}}\right)^{2.95}
\mbox{counts s$^{-1}$ cm$^{-2}$}.
\end{equation}
We note here that the spread about the average at a given $T_x$ 
exhibited in Fig. \ref{fga.fig} is of order of a few. 
This scatter, also pointed out
in \S 3.1, is almost certainly real and is a reflection of the
different peculiar formation history
and current dynamical state that can characterize a group/cluster 
with a given temperature.
With this relation we find that for a Coma-like cluster with temperature 
of $T_x=8.3$ keV the mean $\gamma$-ray flux would be
$\sim 1.4 \times 10^{-8}$ counts s$^{-1}$ cm$^{-2}$. 
Similarly, after rescaling the flux for the
appropriate distance, we can compute $F_\gamma$ expected for other 
clusters of known temperature. 
Thus, we find $F_\gamma \sim 9.8 \times 10^{-9}$ for a temperature
of 6.3 keV and a distance of 55 \hinv Mpc characteristic of 
the Perseus cluster \citep{schwarzetal92};
and $F_\gamma \sim 3.8 \times 10^{-9}$ for
a temperature of 1.8 keV and a distance of
14 \hinv Mpc as estimated for the M87-Virgo system
\citep{bohrin94}.

The values found above for nearby clusters are
well below the limit set by the EGRET experiment of
$4 \times 10^{-8}$ counts s$^{-1}$ \citep{sreeku96}. However,
they should be easily detectable by GLAST, with a sensitivity  
an order of magnitude below the above value.
Our estimate of the $\gamma$-ray fluxes is
somewhat lower than the values computed by other authors
\citep{dasha95,ebkw97},
which are  close to, or slightly in excess
of, the EGRET upper limits.
This comparison is so even after correcting our results for the aforementioned  
systematic underestimate due to resolution effects. 
Differences in  estimates of the $\gamma$-ray flux are 
expected, given the numerous differences in our physical 
assumptions and methodology when compared to the previous authors.
For example, the spatial distribution of CRs given by the simulation
in our case was assumed, by contrast, to be
uniform in \citet[][]{dasha95}, or such as to produce a constant ratio of 
CR to thermal pressure in \citet{ebkw97}.
In addition, we have used a fixed emitting volume within
a radius of 1.3 \hinv Mpc, whereas the previous authors used
$\sim$4 \hinv Mpc for the size of the Coma cluster \citep{dasha95, ebkw97},
and $\sim$1.3 \hinv Mpc for a 
Virgo-like type of cluster \citep{dasha95}.
Also, for the slope of CR {\it energy} distribution
those authors \citep{dasha95,ebkw97}
borrowed the empirical value from the Galactic case; \ie assumed $q=4.7$,
unlike $q=4.0-4.2$ from our simulation. 

Our estimate of the $\gamma$-ray flux for the Coma cluster is,
on the other hand, compatible with the value computed by \citet{cobl98},
although our temperature dependence in Eq. (3.7) is quite a bit steeper than 
the one they presented. 
They computed the emission within the virial radius, 
whereas a fixed volume has been used in our estimate.
In addition, they accounted analytically for a weak phenomenological 
dependence of the baryon fraction on the cluster size,
while the baryon density from numerical simulation
has been used in our estimate.
We note that
the expected functional form of the $\gamma$-ray flux can be modeled as
\begin{equation}
F_\gamma \propto N_{cr}\,n_{b},
\end{equation}
where $N_{cr}$ is the total number of CR 
protons and $n_b$ is the average group/cluster baryonic mass density,
both inside a fixed radius.
So, most of the temperature scaling in our estimate is accounted for
by the following facts: (1) The kinetic energy power available for CRs 
is proportional to $T_x^2$ \citep[see][]{minetal00}. That is,
$E_{cr} \propto T_x^2$. On the other hand,
with a constant momentum slope in the CR distribution function,
$N_{cr} \propto E_{cr}$.
Hence, $N_{cr} \propto T_x^2$ holds approximately in the simulation.
(2) The mean mass density inside a fixed volume (thus, also the associated
baryon mass) scales almost
linearly with the temperature, which is compatible with observations
\citep[\eg][]{edst91b,mohr99}.
Together, those explain the origin of the $\sim T_x^3$ dependence found in the
simulation presented here.

One important point of our findings is that the calculated 
values for the $\gamma$-ray flux are well below the
upper limits set by EGRET, even while a large fraction 
of the total energy in the ICM gas is, in fact, stored in 
CRs. This result differs from
\citet{blasi99}, who finds that the amount of energy in the CR component
must be well below the equipartition value in order not to violate
the same observational limits (except for the extreme
case of a uniform distribution of CRs throughout clusters). 
Blasi's result derives from his adoption of a central
point source for cluster CRs which then must diffuse throughout the
cluster. That leads to a CR distribution more concentrated towards
higher thermal gas densities in the core than if the CRs are
produced by structure shocks as in our case. 
Since in a denser environment the CRs experience many 
more interactions, a higher $\gamma$-ray flux is expected.

\begin{figure}
\plotone{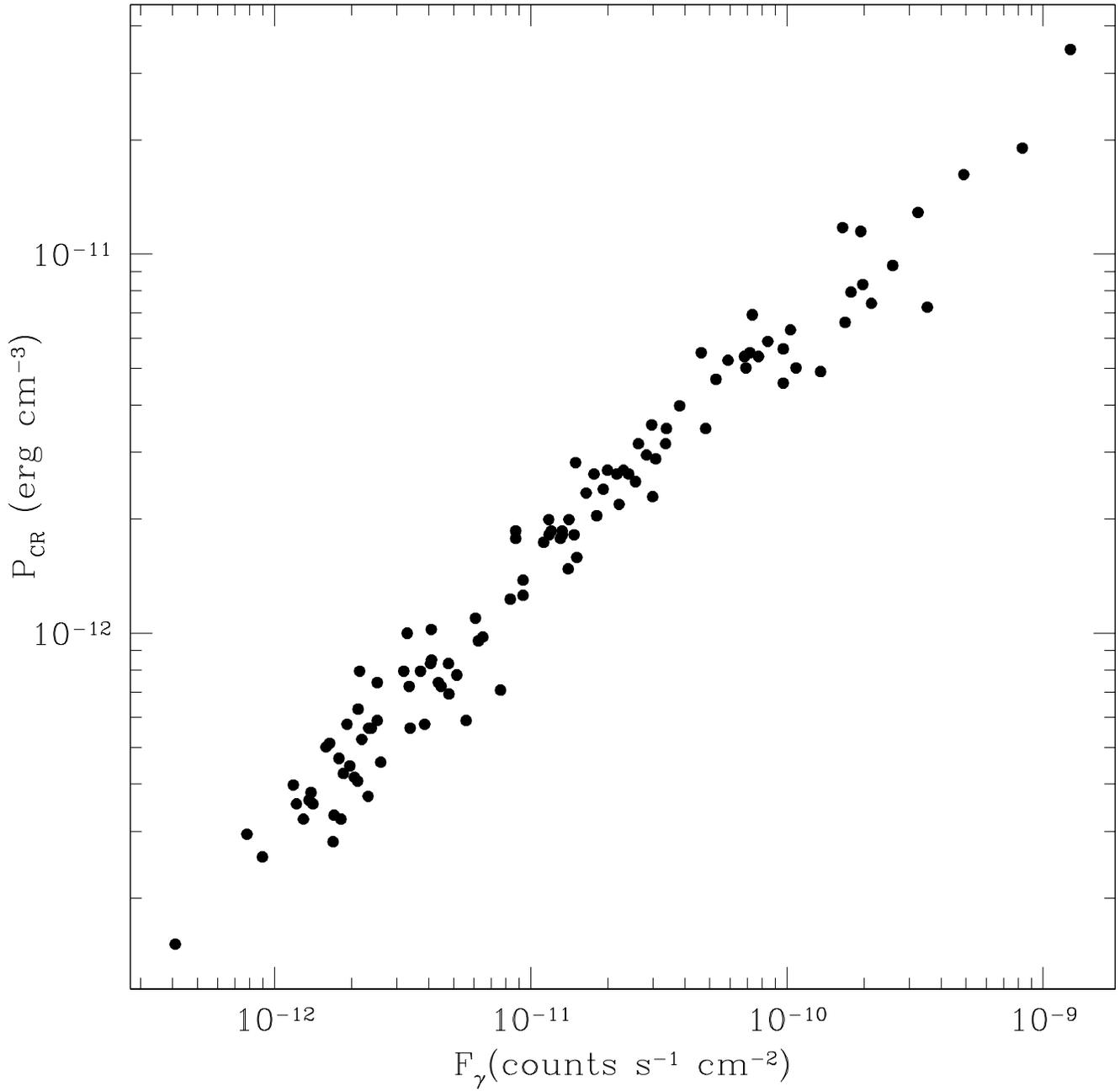}
\figcaption{CR pressure as a function of $\gamma$-ray flux.
\label{fgpcr.fig}}
\end{figure}

Finally, we have found 
that there is a tight correlation
between the CR pressure and the $\gamma$-ray flux that can be fit by 
\begin{equation} \label{pcrfg.eq}
P_{cr}= 2.7\times 10^{-11}\;
\left(\frac{F_\gamma}{10^{-9}
\mbox{counts s$^{-1}$ cm$^{-2}$ }}\right)^{0.64}
\mbox{erg cm$^{-3}$ }.
\end{equation}
This is shown in Fig. \ref{fgpcr.fig}. The above scaling
is compatible with $F_{\gamma} \propto T_x^3$ and
$P_{cr} \propto T_x^2$, if the slope of the CR momentum distribution
varies only slightly (as it is the case here).
$P_{cr} - F_{\gamma}$ is manifestly a cleaner
relation than $P_{cr} - T_x$ (not shown in this paper).
The outcome is not surprising and is due to the direct physical relationship 
between $P_{cr}$ and $F_\gamma$, \ie $F_\gamma \propto \int n_{gas} \, P_{cr}$
(for fixed $q$). Thus the $P_{cr}-F_\gamma$ relation probably allows the
best determination of 
the amount of CR pressure in groups/clusters. 
In addition, measuring the $F_\gamma -T_x$ relation
and comparing it to the numerical results
will provide an important test 
for our numerical treatment and general understanding
of CR injection, transport, and acceleration in group/cluster environment.

\subsection{ X-ray and $\gamma$-ray Images} \label{xgaim.se}

In Fig. \ref{cls1a.fig}, we present synthetic images of
$\gamma-$ray emission from $\pi^0$ decay (left) and
X-ray emission from thermal bremsstrahlung (right)
for a rich group of galaxies in the simulation with $T_x \simeq 3$ keV. 
A larger sample of groups/clusters from the numerical simulation
was studied by \citet{mint00}, but their main properties 
can also be summarized by the findings below.
The volume considered for the realization of the images in
Fig. \ref{cls1a.fig} is a cube
about 14 \hinv Mpc on a side, centered on the collapsed object.
Each synthetic map has a grey-scale bar indicating the logarithmic
value of the imaged quantity. 
The physical units are in `erg cm$^{-2}$ per pixel' for the 
flux density of X-rays, and in `counts cm$^{-2}$ per pixel' for 
the flux density of $\gamma-$rays (see \S \ref{glpr.se} for more details).
In addition to the synthetic images, in Fig. \ref{cls1c.fig} we also present,
for the same object,
two-dimensional slices of the following quantities: 
the CR proton number density in units of cm$^{-3}$ (top left), 
the velocity field (top right), the contours of shock compression 
(${\bf \nabla\cdot v}$ - bottom left) and
the gas number density in units of cm$^{-3}$ (bottom right).
The slices are through the object center and perpendicular 
to the line of sight of the synthetic images. 
The side of the images in each panel of Fig. \ref{cls1c.fig} 
is about 5.3 \hinv Mpc.

First, we note that 
the synthetic $\gamma$-ray image exhibits an irregular
morphology, somewhat different from the smooth slightly ellipsoidal
shape of the thermal X-ray image. This is due to the fact that 
the CRs, responsible for the $\gamma$-ray emissivity, are
sensitive to the particular shock distribution in the ICM.
When even a mild shock crosses through a group/cluster,
the injection of fresh particles over a relatively
short time can significantly enhance the CR population.
This should definitely
affect the shape of the $\gamma$-ray image.
But it is not so for the X-ray image, because the effect of 
a weak shock is only a modest increase of the density (square)
on which the emissivity primarily depends. That effect is
likely to be blended away after line of sight integration
(although finite resolution may limit the amount of 
visible features as well).
In fact, closer inspection by means of two-dimensional 
slice images shows that enhanced
cosmic ray density occurs downstream of
``internal shocks''; \ie post-shock flows
in the central regions of groups/clusters \citep{mint00}.
In Fig. \ref{cls1c.fig} the
vertically elongated, high CR density 
structure (top-left panel) is enclosed by a Mach surface where
a supersonic flow is suddenly decelerated by a shock. 
This region corresponds to a vertical dent in the velocity
vector field (top-right) in the direction N-W from the center
of the panel.
Note that the cosmic rays are where
most of the intra-cluster gas is located (bottom-right panel) in addition being near the shock,
and, therefore, where the injection rate is higher.
Such higher level of structure and more
irregular distribution of
the CRs, as compared to the gas density,
explains in part the high pixel to pixel fluctuation 
of the ratio of CR to thermal pressure that was found
in \S \ref{spdis.se}.

Finally we note that the regions of low surface brightness
corresponding to the same factor below the 
peak value are slightly more extended in the $\gamma$-ray
than in the X-ray image in accord with findings
in \S \ref{spdis.se}. Also,
the highest $\gamma$-ray surface brightness appears more concentrated
than the X-ray brightness distribution. The X-ray emissivity is proportional to the 
square of the gas density, whereas the $\gamma$-ray emissivity is 
proportional to 
the product of the gas density with the CR density.
Thus, this result simply means that the CR protons are 
slightly more concentrated {\it by number}
than the gas in the group/cluster core region.
This result does not contradict the finding in \ref{spdis.se}
that the CR pressure distribution is less concentrated 
than gas pressure. Rather, it indicates that adiabatic 
compression is effective and has reduced the ratio $(P_{cr}/P_{th})$
in the center of the collapsed objects.

\section{Discussion} \label{disc.se}

As we have shown in the previous section, a significant fraction 
of the total energy associated with baryons inside a group/cluster
could be stored in cosmic rays as a consequence of diffusive
particle acceleration at structure formation shocks.
This fact bears important consequences
that we will try to address in the following discussion.

Firstly, if the pressure provided by cosmic rays, $P_{cr} = E_{cr}/3$, 
is large enough, it can affect the dynamics 
of the intra-cluster medium and, therefore, 
both its evolution and equilibrium. 
This is of great concern because groups/clusters
of galaxies are invaluable probes to test cosmological theories
and to measure key cosmological parameters 
\citep[\eg][and references therein]{bops99}.
In fact the present-day abundance (number density) 
of rich clusters of galaxies
sets a strong constraint on 
the total mass content of the universe and the normalization
of the power spectrum of the density perturbation by imposing $\sigma_8 \, 
\Omega_m^{1/2} \simeq 0.5 \pm 0.05 $
\citep{bace92, wef93, ecf96, vili96, pen98}.
In addition, the evolution of the number density of rich clusters
allows one to break the degeneracy of the above result 
and is used to determine both 
$\sigma_8$ and $\Omega_m$ \citep{cmye97, bfc97}.
It is not clear whether and how the evolution of structure would be affected by a 
non-thermal dynamical component. Clearly, since most of the mass is dark,
the growth of the density perturbation would be unchanged for the most part.
However, since the observable universe is made of baryons, 
the specific processes that determine their
dynamical and thermal evolution are of crucial importance,
as they also greatly affect cluster observational properties.
In this respect the effects produced by the cosmic ray 
component could be important, 
even though the underlying large scale structure remains unaffected.

Furthermore, a substantial cosmic ray pressure component could contribute
to the dynamical support of the intra-cluster medium against 
gravitational collapse.
This would affect the 
estimate of the total cluster mass derived from observations and, 
in turn, both of the baryonic 
fraction there and of the total mass of the universe \citep{wnef93}.
Results from a number of studies have
suggested that mass estimates 
based on the hydrostatic equilibrium assumption and X-ray measurements
tend to be somewhat smaller
than those derived from virial estimates and gravitational lensing
\cite[][see also \citealt{mint00} for an extensive discussion
on the issue]{mavi97,hms99,nemafo00,rosabl00,miba95,wu00}.
Part of this could well be the consequence of dynamical effects 
due to the cosmic-ray pressure and also the magnetic field pressure. 
However, the mass discrepancy issue is 
still controversial and the precision of the current measurements,
at the level of 20\% accuracy, 
does not allow a strongly conclusive statement at this point. 
As discussed in the previous section, $\gamma$-ray observations appear
promising in this respect, since according to our prediction
the expected $\gamma$-ray flux from Coma-like clusters 
should be well above
the detection threshold of GLAST \cite[see also][]{blasi99,doen00}.

From the theoretical side, recent numerical simulations have shown that,
in order to construct a viable and realistic depiction of the 
ICM, the various processes taking place there 
need to be accounted for in sufficient detail. 
In particular, it has become clear that the effect of radiative cooling in cluster
cores produces significant quantitative differences 
in numerical simulations in which it is allowed 
\citep[\eg][]{kawhi93,suos98,pcte00,lewis00}.
In particular \citet{suos98} found that cooling can become
catastrophic in the cluster cores unless prevented
by some additional physical processes. 
Similarly \citet{lewis00} found that radiative cooling 
can have dynamical effects on the cluster structure and evolution.
In particular, they concluded that 
the consequences of cooling are global and affect the cluster 
as a whole, despite the fact that
strong cooling is localized in the central region
of a cluster.
According to the study by \citet{lewis00}, however, the catastrophic
character emerging in the Sughinohara \& Ostriker's simulations
is largely inhibited by the feedback of star-formation (gas removal and
heating). Nevertheless, that does not solve the cooling problem completely. 
In fact, the star formation ensuing from the cooling of the
gas produces too large a stellar component (30\% of all the baryons 
instead of the observed 10\% fraction), too high an X-ray luminosity 
by a factor $\sim 3$ and too high a velocity dispersion \citep{lewis00}.
These excesses are driven by a very high
density, stellar dominated core resulting from the effect of cooling. 
In this respect, the presence of a significant 
CR component could reduce the overly 
dramatic effect of radiative cooling and recover some of the observed 
cluster properties, at least in two ways. Firstly, 
CRs provide an additional non-thermal pressure, which
is not dissipated by radiative effects. This hinders the contraction
of the cooling gas; therefore, prolonging the cooling
time and decreasing the rate of conversion of gas into stars.
Secondly, low energy CR ions provide a source of heating that  
tends to balance cooling, once again softening the effect of the latter
\citep{rephaeli77}.
In this respect, CRs are also likely
to affect the dynamics of a cooling flow by means of the two
generic mechanisms described above.

\section{Conclusions} \label{concl.se}

We have carried out a computational study of 
production of CR protons at cosmological shocks 
associated with the large scale structure in 
a SCDM universe.
We have achieved this by carrying out the first
numerical simulation of structure formation 
that includes 
{\it directly} shock acceleration
(in the test-particle limit approximation), transport
and energy losses of the CRs.
CR injection takes place at shocks
according to the thermal leakage prescription, leading to
the injection as CR of a 
fraction about $10^{-4}$ of the thermal protons passing through a 
shock.
According to our results, cosmic ray ions may provide 
a significant fraction of the total pressure in the intra-cluster
medium. The conclusion cannot be made strictly quantitative yet, 
because the complex physics regulating the acceleration 
mechanism cannot be fully simulated, and our simulated group/cluster
structures are still rather coarse. However, we expect 
the CR pressure may account for a few tens of percents of the total 
ICM pressure. The cosmological consequences of this result were addressed
in the previous section. 

A major step forward will be made possible by the advent of 
the next generation of $\gamma$-ray facilities, \ie GLAST.
In fact, we expect $\gamma$-rays will be detected for relatively
nearby massive clusters. That development will probe directly the cosmic ray 
content in clusters of galaxies (see \S \ref{grf.se}).
In addition, 
$\gamma$-ray imaging and spectroscopy will enable us to infer the 
spatial distribution of the CR density and pressure,
once the gas distribution is known (\eg through X-ray data).
That will translate in direct information on the 
the nature of the CR sources.
In fact, if most of the CRs have been expelled by active
galaxies, then their distribution would not be as widespread 
as in the case where the primary sources are cosmic shocks.
This adds to the wealth of critical information
provided by observation  in this band. 
However, most probably 
only the largest clusters and only their innermost regions 
of highest emission will be probed by these instruments, 
owing to the very low surface brightness in the $\gamma$-ray band.
Nevertheless, those detections would still be invaluable
for the study and a much deeper understanding 
of the dynamics of these objects within the next few years.

\acknowledgments

FM wishes to acknowledge a Doctoral Dissertation Fellowship
from the University of Minnesota.
The work of FM and TWJ has been supported in part by NASA grant NAG5-5055,
by NSF grants AST96-16964 and AST00-71167, and by the University of Minnesota
Supercomputing Institute.
DR and HK were supported in part by grant 1999-2-113-001-5 from the
interdisciplinary Research Program of the KOSEF.
We thank Michal Ostrowski, the referee, for useful comments to the paper.
Also we are grateful to T. En\ss lin and S. D. M. White for reading the manuscript and
providing valuable suggestions.
Finally we are grateful to I. L. Tregillis for his ray-tracing code
and to I. V. Moskalenko and  A. W. Strong for 
providing their GALPROP routines.

\clearpage

\bibliographystyle{apj}
\bibliography{papers,books,proceed}


\clearpage 

\clearpage 

\clearpage 

\clearpage 

\clearpage 

\begin{figure}
\figcaption{Synthetic images in 
$\gamma$-rays from $\pi^0$ decay in units counts s$^{-1}$
cm$^{-2}$ per pixel (left) and X-ray from thermal bremsstrahlung 
in units erg s$^{-1}$ cm$^{-2}$ per pixel (right) from a cosmic
volume of (14 \hinv Mpc)$^3$. \label{cls1a.fig}}
\end{figure}

\clearpage 

\begin{figure}
\figcaption{Two-dimensional slice maps of CR proton number density in
units cm$^{-3}$ (top left), velocity field on that plane (top right), 
contours of shock compression (${\bf \nabla\cdot v}$ - bottom left) and
gas number density, again in units cm$^{-3}$ (bottom right).
The side of images is 5.3 \hinv Mpc.
  \label{cls1c.fig}}
\end{figure}

\end{document}

%% file: macros.tex
\def\cf{{cf.~}}
\def\eg{{e.g.,~}}
\def\ie{{i.e.,~}}
\def\lsim{\raise0.3ex\hbox{$<$}\kern-0.75em{\lower0.65ex\hbox{$\sim$}}} 
\def\gsim{\raise0.3ex\hbox{$>$}\kern-0.75em{\lower0.65ex\hbox{$\sim$}}} 
\def\sc1{\raise2.1ex\hbox{\tiny $r\!\!=\!\!4$}\kern-0.95em{\hbox{$=$}}}

\def\cm3{~{\rm cm^{-3}}}

\def\hinv{$h^{-1}$}

\def\euv{extreme ultra-violet~}
\def\hxr{hard X-ray~}
\def\ltsima{$\; \buildrel < \over \sim \;$}
\def\simlt{\lower.5ex\hbox{\ltsima}}
\def\gtsima{$\; \buildrel > \over \sim \;$}
\def\simgt{\lower.5ex\hbox{\gtsima}}

\def\sc{{\rm Science\ }}